# Generative Pretrained Autoregressive Transformer Graph Neural Network applied to the Analysis and Discovery of Novel Proteins

Markus J. Buehler[1,2]*

[1] Laboratory for Atomistic and Molecular Mechanics (LAMM), Massachusetts Institute of Technology, 77 Massachusetts Ave., Cambridge, MA 02139, USA

[2] Center for Computational Science and Engineering, Schwarzman College of Computing, Massachusetts Institute of Technology, 77 Massachusetts Ave., Cambridge, MA 02139, USA

*mbuehler@MIT.EDU

**ABSTRACT**: We report a flexible language-model based deep learning strategy, applied here to solve complex forward and inverse problems in protein modeling, based on an attention neural network that integrates transformer and graph convolutional architectures in a causal multi-headed graph mechanism, to realize a generative pretrained model. The model is applied to predict secondary structure content (per-residue level and overall content), protein solubility, and sequencing tasks. Further trained on inverse tasks, the model is rendered capable of designing proteins with these properties as target features. The model is formulated as a general framework, completely prompt-based, and can be adapted for a variety of downstream tasks. We find that adding additional tasks yields emergent synergies that the model exploits in improving overall performance, beyond what would be possible by training a model on each dataset alone. Case studies are presented to validate the method, yielding protein designs specifically focused on structural proteins, but also exploring the applicability in the design of soluble, antimicrobial biomaterials. While our model is trained to ultimately perform 8 distinct tasks, with available datasets it can be extended to solve additional problems. In a broader sense, this work illustrates a form of multiscale modeling that relates a set of ultimate building blocks (here, byte-level utf8 characters that define the nature of the physical system at hand) to complex output. This materiomic scheme captures complex emergent relationships between universal building block and resulting properties via a synergizing learning capacity to express a set of potentialities embedded in the knowledge used in training, via the interplay of universality and diversity.

**Significance statement**: Predicting the properties of materials based on a flexible description of their structure, environment or process is a long-standing challenge in multiscale modeling. Our MaterioFormer language-model, trained to solve forward and inverse tasks, incorporates a deep learning capacity through attention and graph strategies, to yield a multimodal approach to model and design materials. Since our model is prompt-based and information is encoded consistently via byte-level utf8 tokenization, it can process diverse modalities of information, such as sequence data, description of tasks, and numbers and offers a flexible workflow that integrates human intelligence and AI. Autoregressive training, using pre-training against a large unlabeled dataset, allows for straightforward adjustment of specific objectives.

## 1. Introduction

Multiscale modeling provides a powerful foundation for analysis and design of hierarchical biological materials1–4. Special attention is given to protein materials that form the basis for numerous biological and biologically derived materials5–7. In that realm of analysis, data-driven modeling using machine learning and related approaches has emerged as a powerful strategy8–14 that includes both analysis tasks (such as, predicting properties from sequences) and inverse design tasks (designing proteins or other biomaterials to meet a set of target properties15. Specifically, generative biomaterials science is an emerging frontier in materials discovery and has been applied to proteins16, organic molecules, inorganics including drug design17, bioactive materials18,  and architected materials19–22 among numerous others, recently facilitated by the use of language models23. With the advent of attention-based transformer models in a variety of realizations24–32, we are beginning to see emergent behaviors of such models33 with important questions that should be explored specific to applications in science and engineering, and as explored here, multiscale modeling of biological protein materials.

**Figure 1** shows an overview of the problem tackled in this paper, focused on solving forward and inverse problems (**Figure 1a**). The model features a capacity to both analyze protein sequences within the scope of end-to-end sequence-to-property predictions, as well as generating molecular protein structures to meet a variety of target properties, all within a single model (**Figure 1b**). Our example applications include secondary structure targets (**Table 1**) that are critical for structural, functional and assembly properties of proteins. At the heart of the algorithm used here is a multimodal text-based autoregressive transformer architecture that builds a set of interaction graphs using deep multi-headed attention, which serve as the input for a deep graph convolutional neural network to form a nested transformer-graph architecture (**Figure 2a-b**). This transformer-graph architecture combines an autoregressive, causal self-attention model with a deep graph convolutional neural network (**Figure 2b**). Inputs to the model are completely text-based and allow, through the use of byte-level tokenization (**Figure 2c**) flexible inputs (details see **Materials and Methods**). Due to this formulation, the model can easily be trained, and new tasks be added. By implementing a pretraining strategy we can endow the model with knowledge derived from unlabeled data derived from a variety of diverse protein sequences from across species (details on datasets and training strategy, see **Materials and Methods**).

The plan of this paper is as follows. First, we introduce the overall approach, model development, and validation. We then cover a series of applications studies focused on using the model to design new proteins with targeted properties. We focus on structural proteins, and discuss how this interactive tool can be used to evolve existing proteins (here, silk protein) into new designs, and how it can be used to develop proteins that incorporate antimicrobial motifs into a new protein that has high solubility, and which can also serve as a structural material to achieve multifunctionality. We conclude with a discussion and outlook to future opportunities.

## 2. Results and Discussion

Multi-headed attention mechanisms are used in many existing transformer models in both sequence data and graph data34. The concept to complement the attention mechanism by incorporating a graph convolutional neural network within the causal multi-headed self-attention block yields multi-headed graph-forming convolutional self-attention approach, akin to category theoretical olog models31,35,36. This idea, based on viewing the attention mechanism as a graph-forming framework, allows us to exploit the complex knowledge graphs generated by the attention mechanism, one per each attention head $h$. We do this by performing deep graph convolutional operations on the discovered graphs. Our strategy provides a powerful framework for many other materiomic transformer applications in which we want to exploit attention and graph generation in an integrated, synergistic manner (**Figure 2b**).

With this model, we now proceed to discuss the training strategy and results obtained. We use a multi-stage training strategy to develop a generalizable model that progressively learns first general, then increasingly targeted and complex tasks, as shown in **Figure 3**. Stage I consists of pre-training the model against unlabeled sequences (we explore various pretraining strategies, including one where 15% of the tokens are masked using a corruption strategy, inspired by the training strategy used in BERT29). The use of masking tokens adds additional complexity to the problem by teaching the model not only how to predict the correct next token, but to accomplish this task with missing information about previous sequence elements. Unlike in BERT, here we use causal masking so that the model can only attend to tokens to the left, hence training for both regressive predictions of next tokens and addressing the task under masked circumstances. Since the masking is randomized in each batch, pretraining is less prone to overfitting. We did not, however, find an improvement in the performance while using such masking, and ultimately used the no masking strategy shown in the bottom of **Figure 3b** for the results shown in the paper).

After pretraining (stage I) is complete (for around 70,000 steps) we proceed to stage II (training on forward tasks, see **Table 2** for an overview). **Figure 4** shows the performance of the forward model for the CalculateSS task (predicting overall content of secondary structure, as defined in the Define Secondary Structure of Proteins (DSSP) algorithm, in DSSP 8). Results are shown in **Figure 4a**, and sample secondary structure predictions are shown in **Figure 4b**. The model shows strong forward capacity for a variety of tasks, notably all integrated in a single model. The good performance suggests that there are likely synergy between the training tasks, exploited by a collective and emergent behavior captured by the model.

Next, we explore whether generative tasks can be added to the model, proceeding to training stage III. **Figure 5** shows generative tasks solved, showing examples for generating new proteins based on given ratios of secondary structure content. The designed sequences are shown on the left, images of the folded proteins in the center, and a comparison of the design objective (labeled as GT) with the actually obtained secondary structure content (labeled as Prediction) shown on the right (for DSSP8 and DSSP3, see **Table 1** for definitions of the secondary structure codes).

In stage III the model has also been trained to solve sequence, *i.e*. amino acid residue level, design tasks. **Figure 6** shows results of such sequence-level generative tasks, where the residue-level secondary structure is provided as an input and proteins are designed, The result shows experimentation with design objectives of alpha-helical proteins with varying lengths. A sample task (`regular font`) and output (in **`bold`**) is:

`~Generate<~hhhhhhhhh~>` **`[MSEVAALGVGALDWGKIK]$`**

**Figure 6a** and **b** show results for two different sampling temperatures (**a**, *T*=0.1, **b**, *T*=0.5). For higher sampling temperatures, proteins tend to be more diverse and novel. However, if the temperature increases >1, the design objectives may be less rigorously met. It is noted that the sampling temperature does not refer to real temperature units, rather it is a measure of how much Gaussian noise is added during sampling. A temperature of 1 refers to added noise with a standard deviation of 1, and hence indicates the point where significant effects are expected in terms of influencing the probability distributions of the predictions, and hence the output of the model.

**Figure 7** explores the effect of sampling temperature *T* and sampling threshold (defined as the fraction of highest rated logit candidates from which is sampled from). The higher the temperature, the more diverse the designs become and the less they tend to adhere to the objective. Increasing the sampling threshold and the temperature provides a mechanism to yield highly diverse outcomes.

**Figure 8a** shows the design of a beta-sheet rich protein structure, using the prompt:

`~Generate<~~eeeeee~~eeeeee~~sseeeeess~~~~eeeeee~~eeeeee~~sseeeeess~~~~eeeeee~~eeeeee~~sseeeeess~~>`
**`[MITVTQIQMAGKYTMTITTDADIQQQKGDIMSETLDINDKTLHFVKNVNPANNDMSYELTMSDKVRVVVDGWEGDEVIRKEGHLI]$`**

Similarly, **Figure 8b** shows a design task that yields a combination of a random coil and an alpha-helix. Both tasks are executed well and yield desirable outcomes.

Our training strategy includes solubility prediction and generation tasks, for which we measure the highest accuracy after training stage III. The accuracy of the solubility prediction is 63% (for sequences up to 128 length as tested on the test set reported in [37] and 77% for sequences up to 64 length, for the same test set). While more studies should be done to improve solubility predictions, we can use this trained model to solve forward and inverse solubility tasks, especially for shorter sequences where high accuracy of 77% is found. **Figure S1** shows results for such experiments, designing proteins using a generative solubility task. **Figure S1a** shows two sample proteins designed that are soluble, and **Figure S1b** shows two proteins that are insoluble. All proteins generated are novel and do not yield any hits via a BLAST 38 search. Finally, **Figure S1c** shows how the generative task can be used to re-engineer part of the sequence to render a soluble protein version.

Next, we explore how the model can be used to incorporate multiple functionalities into protein design. **Figure 9** shows such a strategy as applied to alpha-helical antimicrobial peptide design. Starting from 2MWL (amino acid sequence: `VARGWKRKCPLFGKGG`), an antimicrobial peptide proposed in a recent study39. Therein it was shown that this peptide design shows antimicrobial activity against Gram-negative *E. coli* as well as plant pathogens, specifically *X. oryzae* and *X. campestris*. While the original peptide is unstructured (**Figure 9b**, top), we seek to develop sequences that include the motif VARGWKRKCPLFGKGG but that yields an alpha-helix rich design that will likely help toward the assembly of structural materials and films. To do this we use the Generate task:

Generate<~hhhhhhhhhhhhhhhhhhhhhhhhhhhhhhhhhhhhhhhhhhhhhhhhhhhh~>
[VARGWKRKCPLFGKGG).

We repeat sampling multiple times and assess designs against structural properties and solubility, to create a set of possible designs that can be screened for performance. **Figure 9a** shows an overview of the various metrics used, shown over 10 sampling processes. Then, **Figure 9b** shows visual representations of the predicted candidate proteins. The best performing candidate is sample number 9, being a peptide to be predicted to be soluble with the highest alpha-helix content. Sampling was conducted with *T*=.5, and filter_thres=0.9. This repeated sampling also shows that the model can reliably predict proteins at the desired length (length cue is given by the secondary structure specification in the Generate prompt. Moreover, since our model shows good accuracy for solubility predictions for short sequences, we have additional confidence that the solubility screening has reasonable accuracy.

The flexible approach by which tasks are used, and where one output can be used to construct another, is illustrated in the next example. **Figure 10** depicts the results of a series of experiments using an existing protein, Sericin 1 (*bombyx mori*, P07856 SERI1_BOMMO), and re-engineering the natural protein towards distinct design objectives. Panel **Figure 10a** shows the original proteins structure and sequence of sericin. **Figure 10b** shows a sequence completion task, where the initial sequence is continued in an unconstrained manner. **Figure 10c** shows a design task where the design objective is provided alongside the original sequence and then continued to meet the design task. The design task in this case is to generate an alpha-helical protein, which is indeed found towards the end of the protein. **Figure 10d** shows a similar example, however, with the design task to generate a beta-sheet rich protein. **Figure 10e** shows another example where the design task is given is a target with 50% beta-sheet, 20 random coil content. This design task results in a more complex protein structure, showing that the model has the capacity to profoundly reconstruct an incipient sequence.

As another example, **Figure 11** depicts results of experiments using an unstructured protein sequence designed earlier (see **Figure 5**, bottom example). We expand on this earlier, novel peptide design it using residue-level secondary structure design. As can be seen in **Figure 11a**, the random-coil sequence GYVLGS can be transformed into a beta-sheet rich structure. Similarly, the original design can be re-engineered to form an alpha-helix rich protein. Based on a totally different framing of the task, **Figures 11c** and **d** show experiments where we use two naturally occurring proteins, vimentin 3GE1 and amyloid-forming peptide 2ONV, and query the algorithm to create an alpha-helix rich product. Similarly, in panel **Figure 11d** we show an experiment where we use this combined sequence and query the algorithm to continue the sequence using the Sequence task. This results in an alpha-helix rich structure as well. Such experiments, along with proper scoring functions to assess properties, can be a powerful tool to explore the wider proteome for new designs with applications as drugs, biomaterials, coatings, and others.

Finally, to examine whether the framework developed here can be used to achieve higher accuracy for protein solubility predictions we conducted several experiments and achieved a solubility accuracy of 74% for the test set reported in 37], and 78% for the eSol test set reported in [40. This model is based on a larger pre-training reservoir using UniRef50, for sequences up to 512 lengths (the model is larger with 24 layer depth, a dimension of 1024, and no graph neural net layers for simplicity; training is conducted using layer-wise learning rate decay 41 (LLRD, a method in which we use higher learning rates for top layers and lower learning rates for bottom layers).

## 3. Conclusions

We have shown that generative language methods provide a flexible platform for protein materials discovery and design. We can easily incorporate these models into a wide range of applications and solve multiple, complex tasks, as summarized in **Table 2**. While we have considered a total of 8 tasks in this work, these can be easily extended to feature additional tasks, which provides more data for the model to learn on. While our model solves these tasks overall well, there are certain advantages of using dedicated models that focus on one task at a time (e.g. sequence-to-property predictions, or generative tasks using diffusion models16). For instance, in the design task to create protein sequences that meet a certain per-residue secondary structure, the model reported in this paper sometimes fails to accurately reflect the desired length in the prediction. A similar aspect is seen when secondary structure predictions are made from an input protein sequence. In contrast, a diffusion model trained against solely one generative task 16 solves it more accurately when it comes to the sequence length. However, it is noted that the model in 16 that generated sequences from overall secondary structure contents struggled to identify novel protein designs. The model reported here can solve this task exceptionally well, with a very high degree of novel protein sequence designs.

An appealing aspect of the MaterioFormer model is the flexible, iterative workflow that can integrate human intelligence and AI. As done in the various examples shown (**Figures 5-11**), humans can enter a prompt, design a protein and check whether it suits the design criteria (and if not resample or adapt the design parameters) and then use the output in a secondary task. This was demonstrated in **Figure 11** where we used an initial novel peptide design obtained in **Figure 5**, as well as via the amalgamation of two naturally occurring sequences that however never occur jointly in a protein. Such iterative processes can also easily be combined with autonomous experimentation, providing an additional source of data generation, collection, and further training the model.

On a more theoretical side, the problem solved here is a complex building block assembly problem – building blocks are not just amino acid residues, secondary structures, but also numbers and various tasks by which these numerous combinatorial spaces are combined. Remarkably, the strategy used here learns foundational and transferrable insights. This results, as shown here, in a remarkable wealth of conditioned protein designs as well as forward and inverse task solution. With more data, it is anticipated that highly complex phenomena can be captured.

While the secondary structure predictions are generally good, and especially for the overall secondary structure ratios, the accuracy of solubility predictions remains relatively low compared with dedicated solubility models. However, the accuracy reaches 0.77 for short sequences <64 residues, which represents a good performance. Overall, one could argue that this is a remarkable performance since this task was trained only on a small set of ~4,000 sequence-solubility pairs with proteins of <128 length (as opposed to 40,000 sequences in the whole dataset of sequences with all lengths up to ~1,700). With a deeper model and more pretraining, solubility accuracy reaches up to 78% for sequences up to 512 amino acid lengths, showing great potential for the approach developed here to expand the usability, accuracy and generalizability. Future work could expand the training task of the model to consider even longer tasks and predictions.

The training strategy used here, comprised of text-based prompts, is flexible and can easily be adapted to a variety of tasks. Moreover, since we train and predict numbers encoded as text, we do not have to specially encode numerical values specifically (however, this can be done easily and would allow models to be trained to deal with very high-dimensional data, e.g. fields, images, time-series, which can be easily accomplished using vector-discretized encoding methods such as discrete variational autoencoder as done in [42]). This can be helpful for both task and prediction development, and can allow for encapsulation of high-dimensional data within the architecture. There are also opportunities to introduce cross-attention mechanisms for more complex amalgamation of information processed in the attention and graph layers.

Other future explorations could incorporate additional prediction tasks in both forward and inverse directions, and expand the training set to incorporate more sequences (e.g. during the pretraining stage). It would also be interesting to explore interactions with distinct biological molecules, such as mRNA or DNA, which can be added to the task training due to the flexible byte-level tokenizer. Such training tasks may also feature multiple-scale questions, such as coding not only the constituting proteins or biomolecules, but also other features such as relative concentrations, pH or salt concentration, and others. This may ultimately be used to construct multi-modal multi-scale models that can incorporate knowledge developed from disparate simulation and experimental paradigms into all stages of training, from pretraining to tasks. A multiscale scheme as used in this study captures complex emergent relationships between the basic building block of matter and resulting properties. Hence, it offers a synergizing learning capacity to express a set of potentialities embedded in the foundational knowledge used to train the model that exploits unknown or little understood cross-fertilizing relationships.

Mechanistically, this is facilitated by the elementary design of the approach to use a set of universal building blocks arranged in complex hierarchical patters to create emergent functions [43–45].

## 4. Materials and Methods

### *4.1 Dataset construction and tokenizer*

Pretraining (**Figure 3**, stage I) is conducted with a dataset of ~333,000 sequences collected from the AlphaFold2 prediction database from a variety of organisms including *Human, M. jannaschii, Mouse, Maiz,* and many others (https://alphafold.ebi.ac.uk/; sequences up to a length of 256 amino acids are used, constructed from sequences for UP000000805, UP000008816, UP000001584, UP000000625, UP000002485, UP000001450,UP000000559, UP000002311, UP000008153, UP000002195, UP000000803, UP000002296, UP000001940, UP000005640, UP000002494, UP000000589, UP000000437, UP000006548, UP000008827, UP000007305, UP000059680, and Swiss-Prot). For secondary structure tasks, we use the dataset reported in 46 that consists of 125,000 sequences for which overall secondary structure content and per-residue DSSP predictions have been calculated (of these we select sequences with 128 or less amino acids, or a total of ~14,500 sequences, for training). Solubility is trained on the dataset reported in 37 with 40,000 sequences with associated solubility label (0 or 1) (we select sequences of 128 or less amino acids, resulting in ~4,300 sequences for training and 293 test sequences; from "Test Set 1" in 37). In the larger model we use around 37,000 sequences of up to 512 length for training, and 1792 sequences for testing solubility predictions.

We use byte-level tokenization to represent UCS Transformation Format 8 (utf8) codes in 256 tokens (in this scheme, each character is represented by one to four bytes). This strategy allows us to encode the nature of the physical system at hand, including a variety of tasks, and also opens the door to future training/fine-tuning of the model to include other sequences (e.g. amino acids, variants of natural amino acids, DNA), tasks, or chemistries (e.g. SMILES). The distributions of tokens as obtained for the training set used in this study is shown in **Figure 2b**. Token sequences are encoded using trainable embedding layers. A physical interpretation of this strategy is that it defines all aspects of the physical including any operation applied to it, such as a design task or calculating properties. In a more abstract sense, it defines the system and tasks from an elementary building block perspective that can include both numbers, characters, symbols, or other features. The model then learns to understand the relationship between these building blocks in order to solve tasks.

### *4.2 Attention based transformer and graph-convolutional models*

**Figure 2a-b** depicts a summary of the autoregressive transformer architecture, representing a decoder-only architecture that produces solutions iteratively from a start token during inference. The key mathematical operation is the masked attention mechanism23,32, defined as

$$\text{Attention}(Q, K, V; M) = \text{softmax}\left(\frac{QK^T + M}{\sqrt{d_k}}\right)V \tag{1}$$

with a triangular mask $M$ (here, for a sequence of length 3):

$$M = \begin{pmatrix} 0 & -\infty & -\infty \\ 0 & 0 & -\infty \\ 0 & 0 & 0 \end{pmatrix} \tag{2}$$

(so that the model can only attend to tokens to the left (*i.e.*, previous tokens to enforce causality). The causal attention calculation is implemented in multi-headed form by using parallelly stacked attention layers, with a total dimension $d$. Instead of only computing the attention once, in the multi-head strategy (where $h$ denotes the number of attention heads) we divide the input into segments (in the dimension of the hidden dimension, that is, $d_{v,i} = d/h$). We compute the scaled dot-product attention over each segment, allowing the model to jointly attend to information from different representation subspaces at different positions:

$$\text{MultiHead}(Q, K, V) = \text{Concat}(\text{head}_1, \ldots, \text{head}_h)W^O \tag{3a}$$

$$\text{head}_i = \text{Attention}\left(QW_i^Q, KW_i^K, VW_i^V\right) \quad (3b$$

In self-attention as used in this work, all $Q$, $K$, $V$ come from either input or output embeddings (or other sources) only.

We complement this standard transformer architecture by constructing a per-head graph neural network architecture based on the attention scores, providing edge features. Based on the approach shown in **Figure 1b**,

$$E = \text{softmax}\left((QK^T + M)/\sqrt{d_k}\right) \quad (4)$$

is used to define directed edge features of a set of $h$ graphs $\mathcal{G}_i$, each of which with $N$ nodes ($N$ being the length of the input sequence), where $E_{ij}$ defines the edge feature of node $i$ to $j$ ($E \in \mathbb{R}^{N\times N}$). Node features of the graph are defined by the corresponding part of $V$, or

$$V_i \ = VW_i^V \ (\in \mathbb{R}^{N\times d_{v,i}}). \quad (5)$$

A series of graph convolutional operators are applied by creating a deep graph neural network with $N_{\text{GNN}}$ layers (in our implementation, the hidden dimension is equal to each head dimension, $d_{v,i}$, but this could be changed in principle to allow for additional learning capacity). We use message passing to the neighbors defined by all non-zero elements in $E$, and use mean aggregation weighted by all edge features to update node features. Graph processing is conducted for each of the $h$ graphs, and the resulting node features $V_i$ are then concatenated to form $V$. This way, the output of the graph convolutional processing and the scaled dot product attention have the same dimension, $\mathbb{R}^{N\times d_v}$. Since $M$ is a triangulated causal mask used in the construction of $E$, causality is retained in the graph convolutional operators via a directed graph.

The result of the deep convolutional graph neural network and the regular multi-headed attention operation are combined additively. Gaussian error linear unit (GELU) activation functions 47 are used in both the transformer and graph convolutional neural structures.

A start token $\mathcal{T} = \sim$ is added at the beginning of the prompt, so that

$$z = [\mathcal{T}, z_1, z_2, \ldots, z_N] \quad (6)$$

During generation, the start token $\mathcal{T}$ followed by the task is fed into the model and the output is predicted from it. During sampling iterations, this process is repeated until the full output is produced, capped using an end token $\mathcal{E} = \$$. All conditioning and distinction of various tasks is provided by the input prompt. Additional sets of tokens are defined to encapsulate various tasks and input/output boundaries <..> encapsulate task, [..] to encapsulate prediction). Causal autoregressive training is performed using cross-entropy loss, where the next token in the input sequence $z$ is the label for the current token (*i.e.*, labels start with the second token of the input, and we remove the last logit since no label exists, see **Figure 3b**). The training data consists of $\mathcal{T}$, followed by the task and the corresponding prediction, ending with $\mathcal{E}$:

$$z = [\mathcal{T}, z_1, z_2, \ldots, z_N, \mathcal{E}] \quad (7)$$

Gumbel softmax sampling48,49 is used during inference. This allows one to adapt the creativity of the model. This is achieved by adding a defined level of noise controlled via the sampling temperature $T$ to a fractional set of logit distributions (identified by a sampling threshold) predicted by the transformer model. We then sample the predicted token from this revised distribution. This helps to add expressivity to the generative tasks to achieve more variations in the predictions ($T$ around or larger than 1). In forward prediction tasks we find that they are best conducted using low sampling temperatures ($T$=0.1 or lower). The model features a dimension of 256, 8 heads (each with $d_{v,i}$ =32 dimension), depth=12, feed forward multiplier =4 (1024 channels), dropout=0.1, embedding dimension=32, 3 GNN layers nested within each of the 12 transformer layers (the hidden dimension of the GNN is 32, same as the head dimension $d_{v,i}$). Positional encoding is realized via

Fourier encoding. The total depth of the model features 12 (transformer decoder layers) x 3 (GNN layers)=36 total layers.

The predicted sequences are folded into 3D protein structures using OmegaFold50, and further analyzed using DSSP 51,52 (to obtain secondary structure information). Additional analysis to assess the novelty of sequences generated is conducted using BLAST[38].

**Table 2** summarizes all prompts used in the model.

### *4.2 Training process and other hyperparameters*

All code is developed in PyTorch 53. All machine learning training is performed using an Adam optimizer 54, with a learning rate of 0.0002. We use between 2000 and 4000 warmup training steps (during which the learning rate is ramped from 0 to the desired learning rate), followed by exponential decay.

**Figure 3** depicts the training strategy, featuring a total of three stages. The first stage represents general-purpose pretraining. We use both masked (15% of the input tokens are randomly masked with a masking token "_") and unmasked pretraining. We find that the unmasked pretraining strategy yielded overall better results, but it deserves further exploration and may be advantageous in certain scenarios. For instance, we found that masked pretraining yielded a better performance in certain forward tasks. The second stage focuses on training forward tasks (calculating various protein properties), and the third stage trains on both forward and inverse tasks (designing sequences to meet a certain target). Fewer training epochs are needed from left to right, as the model learns complex relationships and ultimately synergistically builds on knowledge from forward and inverse tasks.

## Author contributions

M.J.B. developed the overall concept and the algorithm, designed the ML model, developed the codes, oversaw the work, and drafted the paper.

**Code availability**: The MateriomicTransformer code, trained weights, and data is available at: https://github.com/lamm-mit/MateriomicTransformer.

**Supplementary Material**: Includes Supplementary **Figure S1**, designing proteins using a generative solubility task.

**Acknowledgements**: This work was supported by the MIT-IBM Watson AI Lab, the Army Research Office (W911NF1920098 & W911NF2220213), ONR (N00014-19-1-2375 and N00014-20-1-2189), as well as USDA (2021-69012-35978).

[54] D.P. Kingma, and J. Ba, “Adam: A Method for Stochastic Optimization,” https://arxiv.org/abs/1412.6980 (2014).

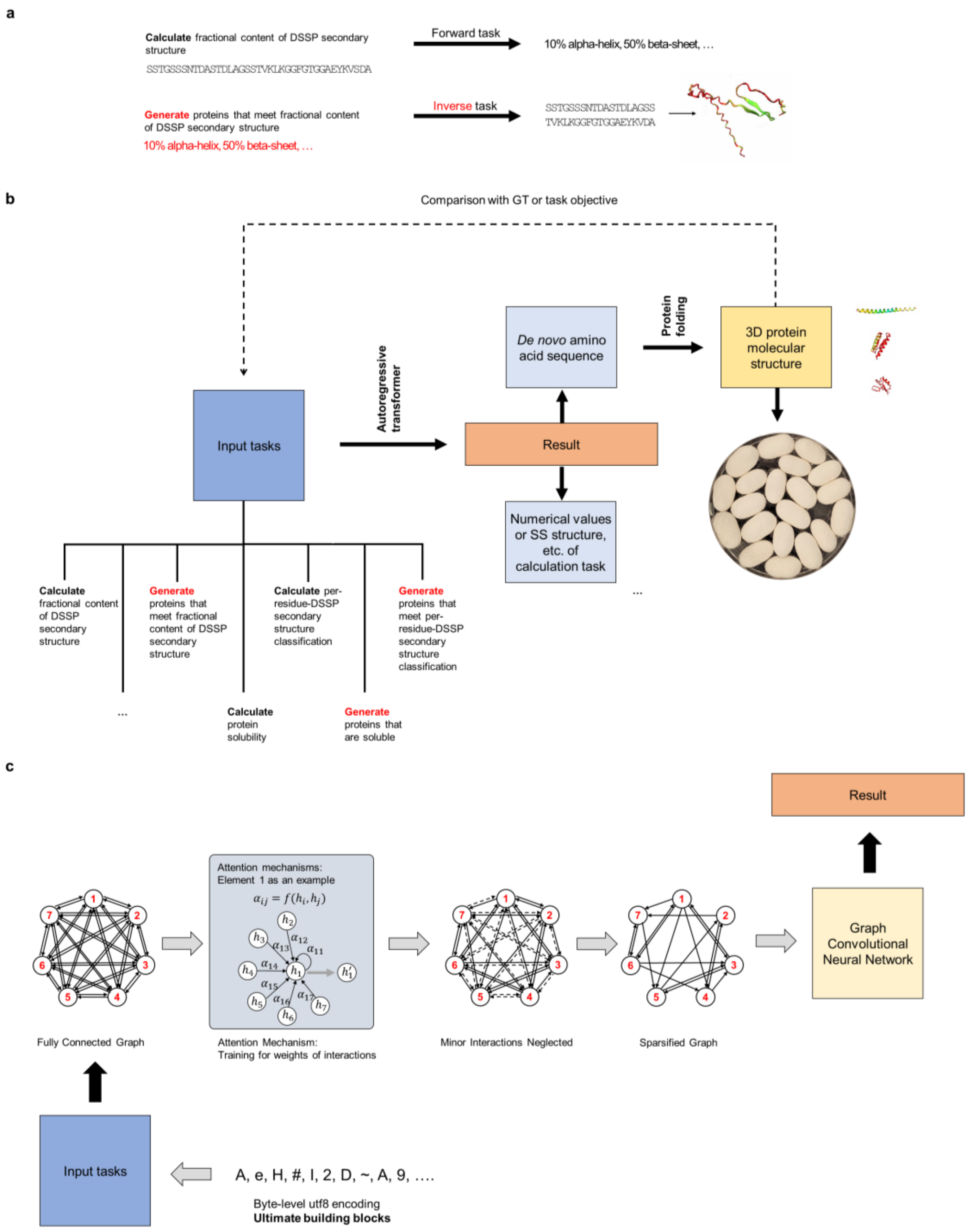

**Figure 1**: A deep language model is developed that can solve forward and inverse protein modeling problems. Panel **a** shows two sample tasks, forward (e.g. calculate secondary structure content of a protein given its sequence) and inverse (design a protein to meet a specified secondary structure content). Overview of the approach implemented, generating molecular structures from amino acid sequences (panel **b**). The model realizes a variety of calculate and generate tasks to solve multiple protein analysis and design problems. At the heart of the algorithm used here is a text-based transformer architecture that builds interaction graphs using deep multi-headed attention, which serve as the input for a deep graph convolutional neural network to form a nested transformer-graph architecture (panel **c**). In a broader sense, the modeling conducted here relates an ultimate set of building blocks – here, byte-level utf8 encoded characters – to complex output, which can take many forms. This multiscale scheme captures complex emergent relationships between the basic building block of matter and resulting properties. DSSP is the acronym that refers to the Define Secondary Structure of Proteins (DSSP) algorithm.

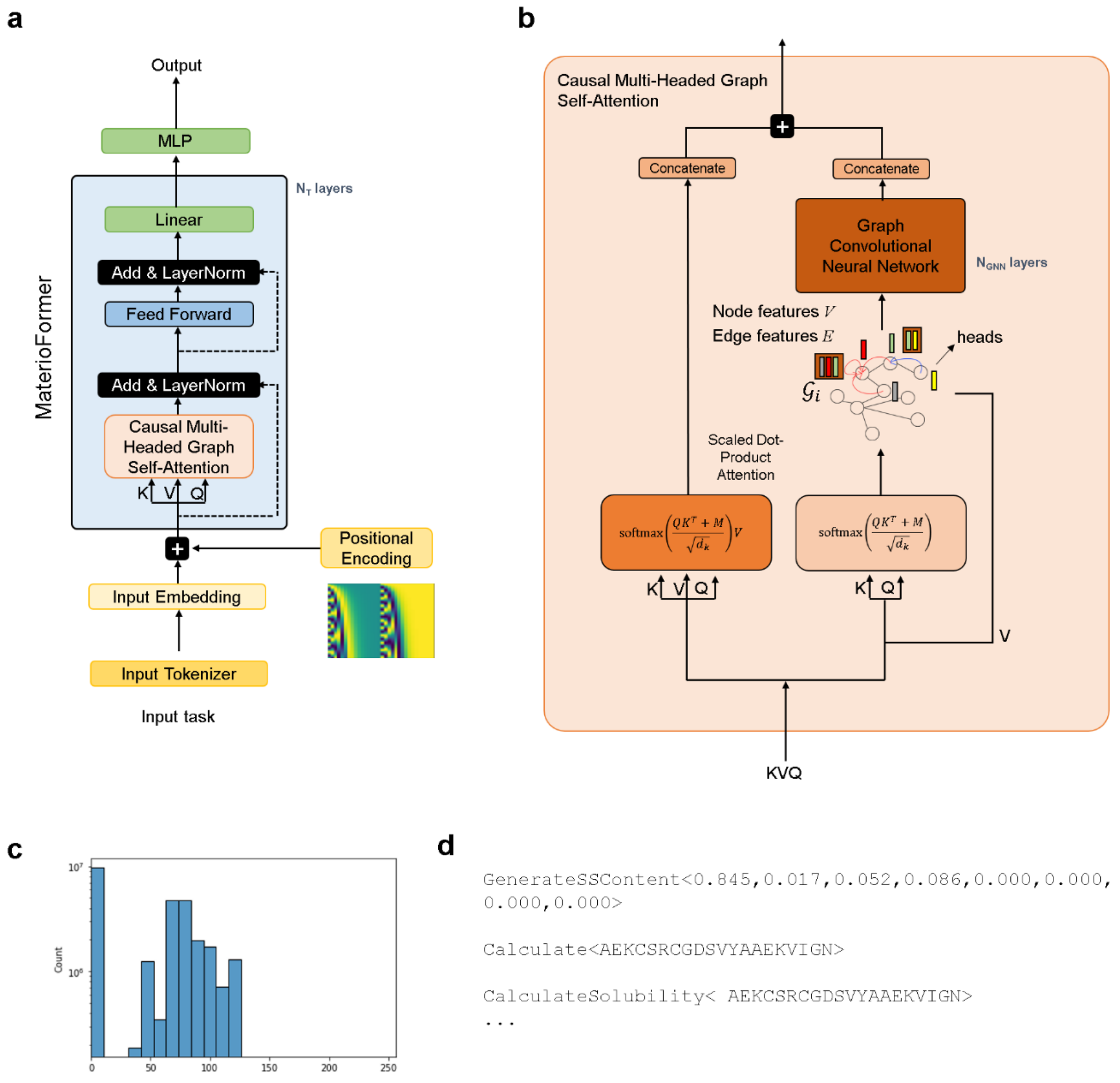

**Figure 2**: Overview of the MaterioFormer model, an autoregressive transformer-graph convolutional model built on text-based prompt input for diverse tasks. Panel **a** depicts details of the implementation of the model, with **b** showing the causal multi-headed graph self-attention strategy used. The model features a conventional scaled dot-product attention mechanism, using causal self-attention via the triangular mask M, complemented by a graph convolutional neural network. Based on the concept schematically shown in **Figure 1b**, $\text{softmax}\left((QK^T + M)/\sqrt{d_k}\right)$ is used to define the edge features of a set of $N_{\text{heads}}$ graphs, each of which with $N$ nodes ($N$ being the length of the input sequence), with node features defined by the corresponding part of $V$. Graph convolutional operators are applied by creating a deep graph neural network with $N_{\text{GNN}}$ layers (hidden dimension is equal to each head dimension, generally $d_{\text{head}} = d/N_{\text{heads}}$). The message passing approach is schematically illustrated in the image, where edge information is used to scale aggregation. Panel **c** shows the statistics of the byte tokenizer, which encodes generic utf8 string data into 256 distinct tokens. The padding token is 0, most commonly seen in the padded sequence data fed to the model. Panel **d** shows sample prompts used (a complete list, see **Table 2**).

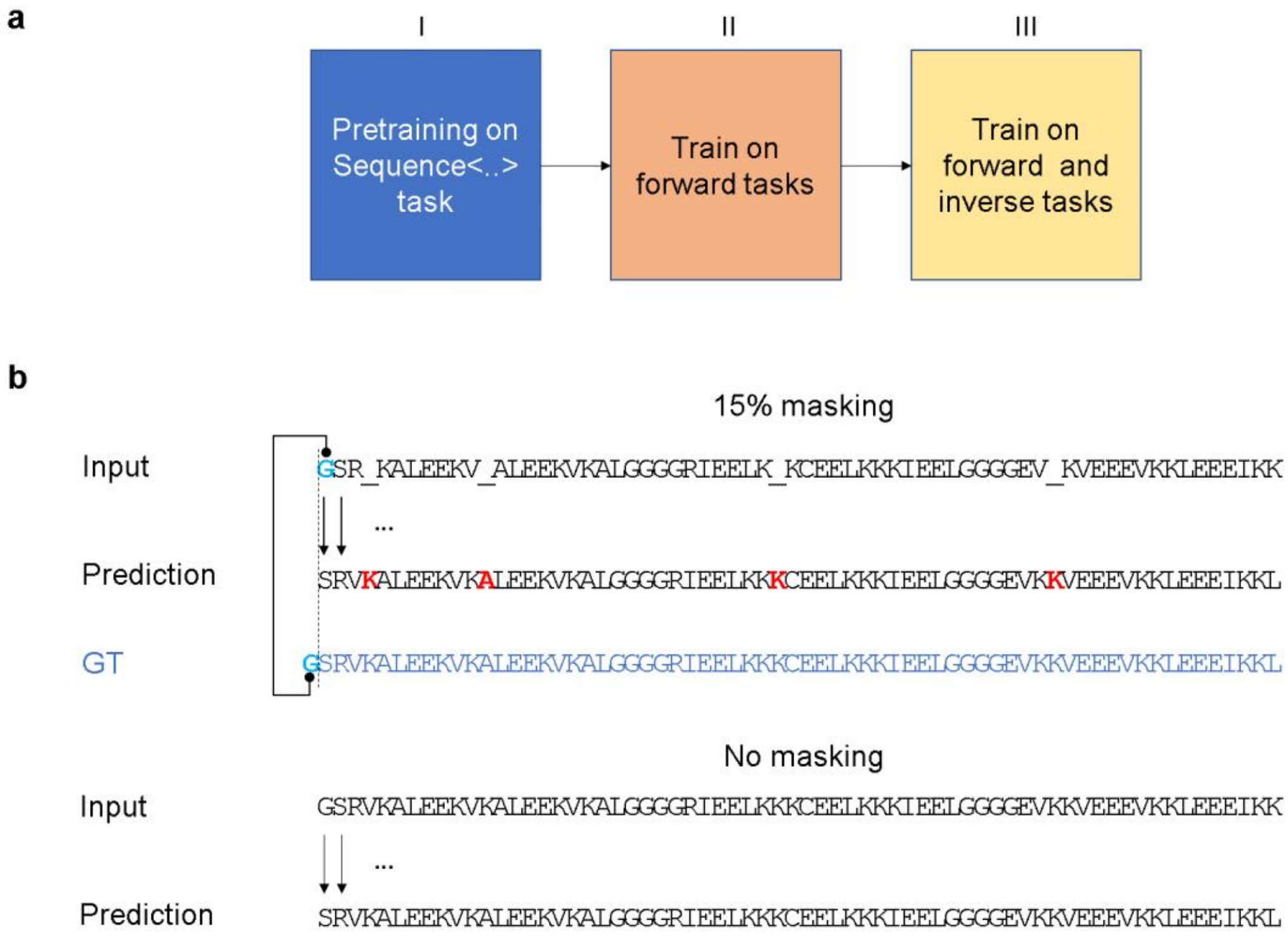

**Figure 3**: Training strategy, featuring three stages (**a**). The first stage represents general-purpose masked pretraining (as shown in **b** we explore both, a strategy where we corrupt 15% of the input tokens randomly [randomized every in every training step] with a masking token "_"], and a pretraining strategy without masking). We use ~333,000 unlabeled sequences as training data. The second stage focuses on training forward tasks (calculating various protein properties), and the third stage trains on both forward and inverse tasks (designing sequences to meet a certain target). Fewer training epochs are needed from left to right, as the model learns complex relationships and ultimately synergistically builds on knowledge from forward and inverse tasks.

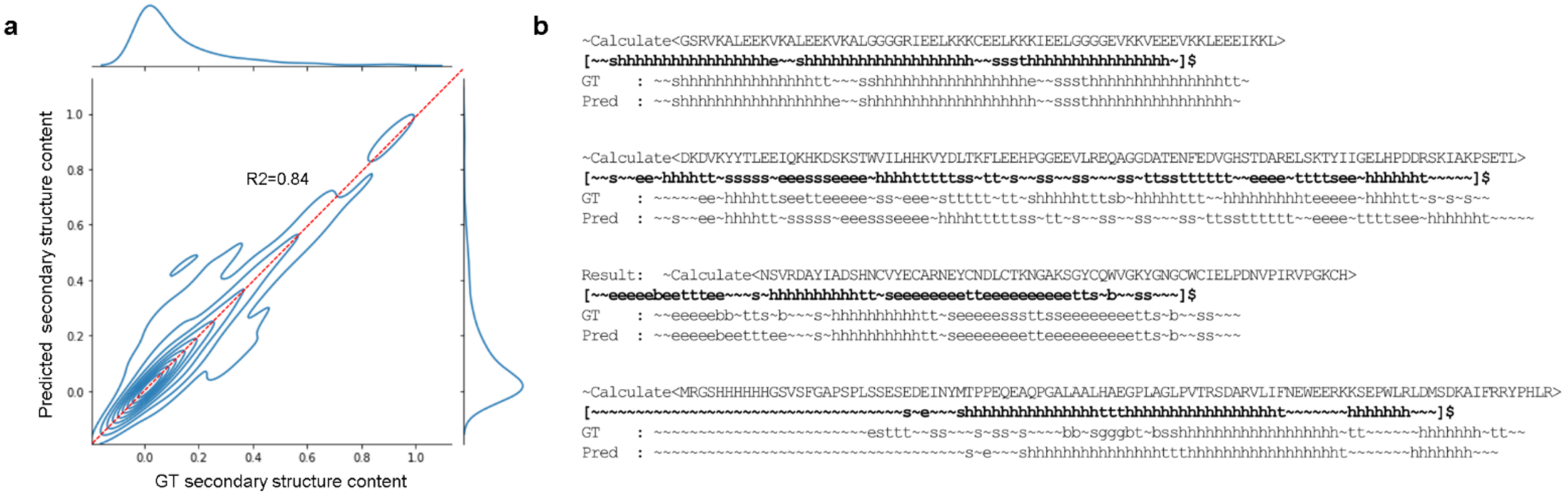

**Figure 4**: Performance of the forward model after training stage II, for the CalculateSS task (predicting overall content of secondary structure in DSSP 8), depicted in panel **a**. Sample secondary structure predictions are shown in panel **b**. See **Table 1** for a definition of the secondary structure symbols (s, h, etc.).

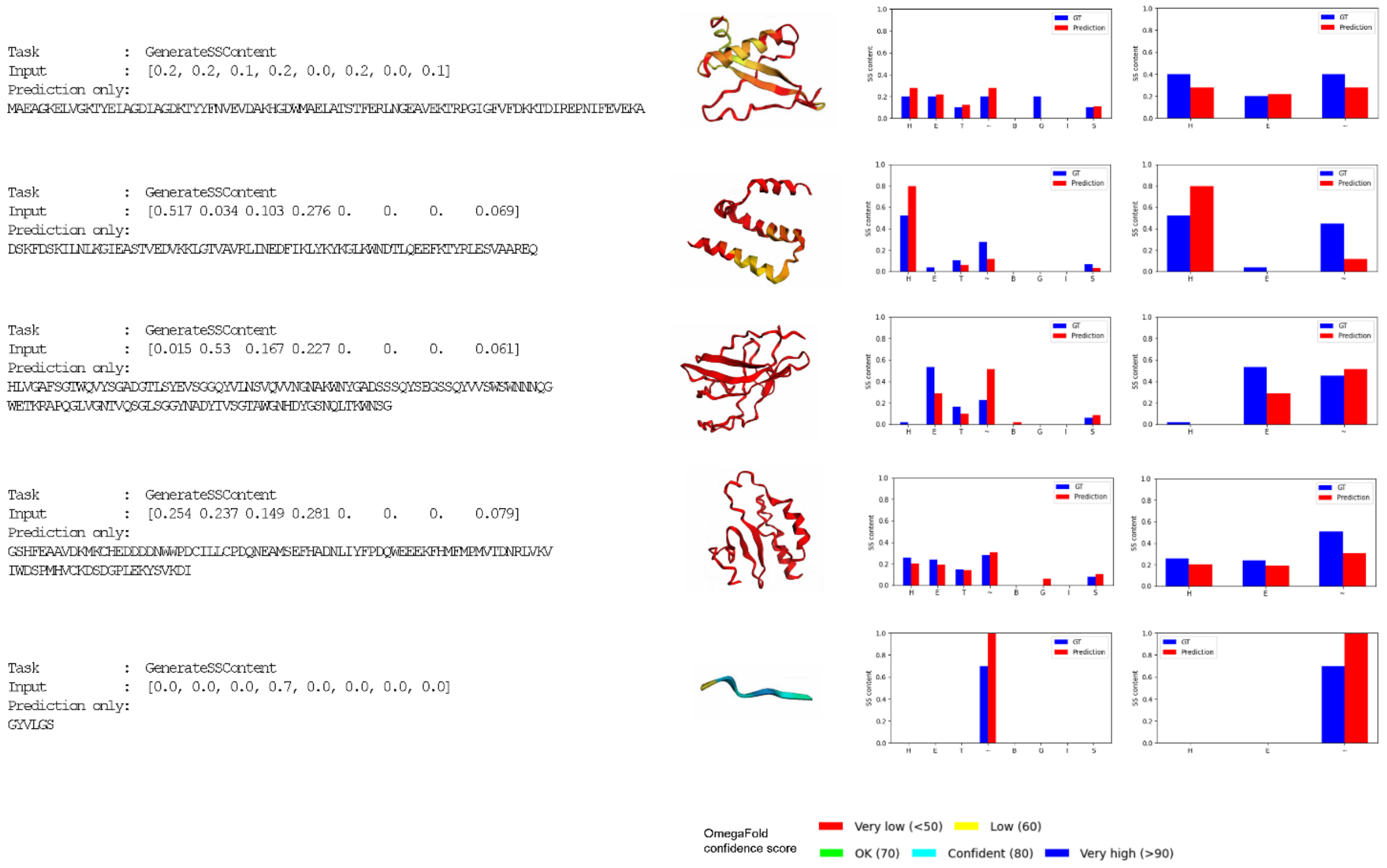

**Figure 5**: Generative tasks solved after training stage III (see **Figure 3** for an overview), showing examples for generating new proteins based on given ratios of secondary structure content. The designed sequences are shown on the left, images of the folded proteins in the center, and a comparison of the design objective (labeled as GT) with the actually obtained secondary structure content (Prediction) shown on the right (for DSSP8 and DSSP3, see **Table 1** for definitions). All proteins visualized in this paper are colored (per residue) by the confidence score 50).

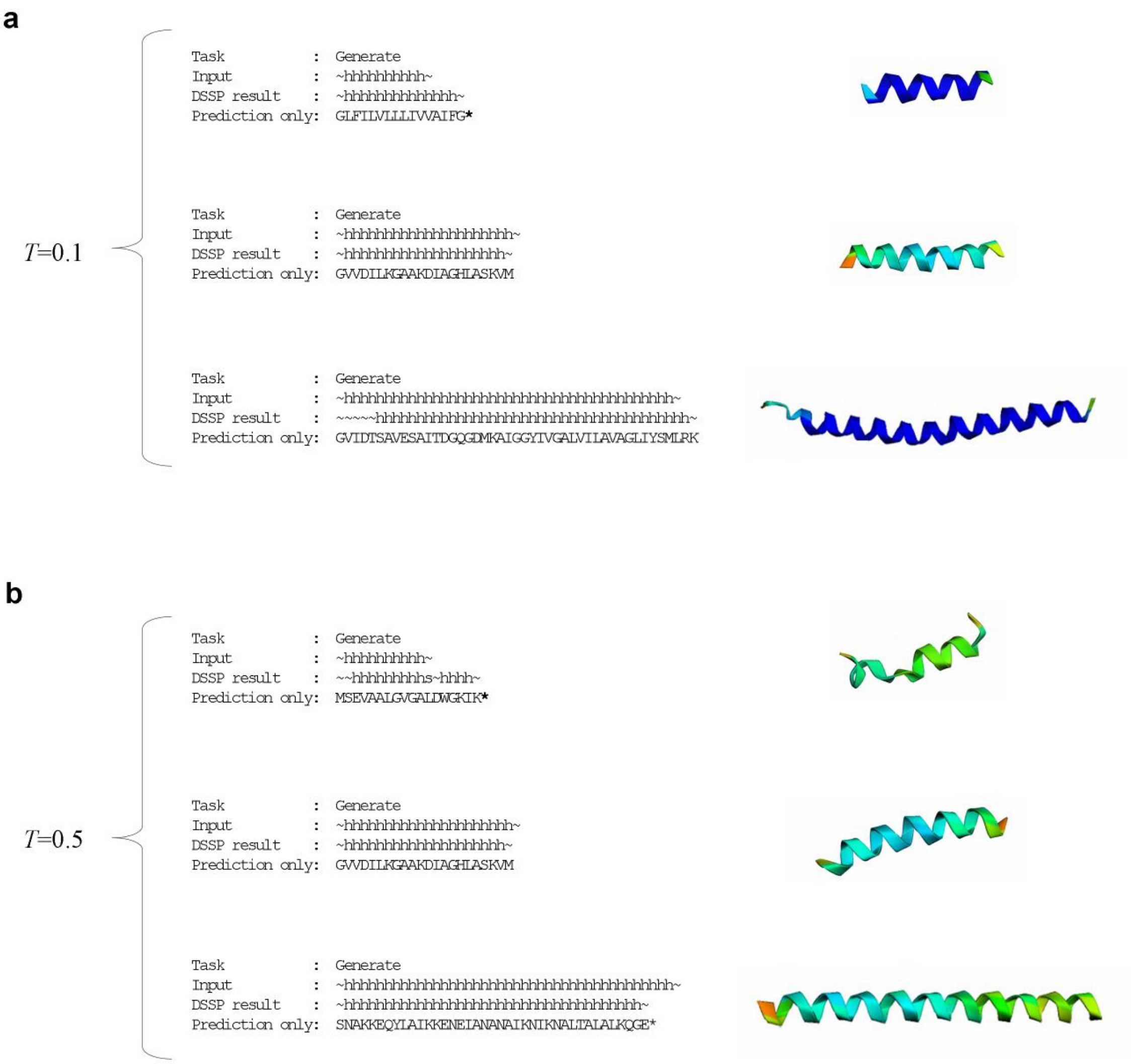

**Figure 6**: Sequence-level generative tasks, where the residue-level secondary structure is provided as an input and proteins are designed, The result shows experimentation with design objectives of alpha-helical proteins with varying lengths. A sample task and output is ~Generate<~hhhhhhhhhh~> **[MSEVAALGVGALDWGKIK]$.** Panels **a** and **b** show results for different sampling temperatures (a, *T*=0.1, b, *T*=0.5). For higher sampling temperatures, proteins tend to be more diverse and novel but if the temperature increases >1, the design objectives may be less rigorously met. Sequences marked with * are novel.

| | |
|---|---|
| | Task : Generate<br>Input : ~~hhhhhhhhhhh~ |
| *T*=0.25<br>thresh=0.9 | DSSP result : ~hhhhhhhhhhhh~<br>Prediction only: GLFGAIAGIIEKAI |
| *T*=1.5<br>thresh=0.9 | DSSP result : ~~hhhhhhs~hhhhhhhhhhhs~ggggg~<br>Prediction only: YTAERIQERYPSAAILQWQFSKPQFSSML |
| *T*=3.5<br>thresh=0.8 | DSSP result : ~~~~~~~~~eeeeetteeeee~<br>Prediction only: HFVRDQIDYVCVKTDGKTHKRY |

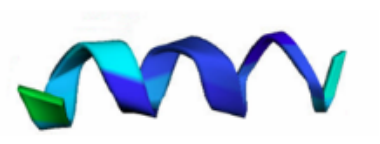

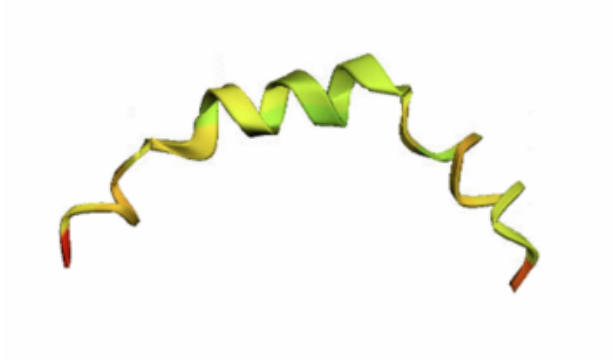

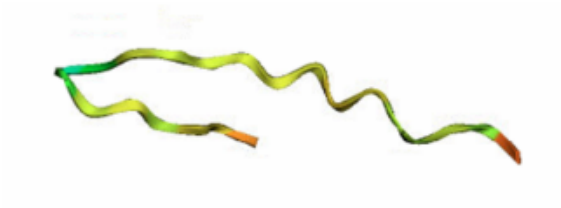

**Figure 7**: Effect of sampling temperature T and threshold. The higher the temperature, the more diverse the designs become and the less they tend to adhere to the objective. Increasing the threshold (defines the fraction of highest rated logit candidates from which is sampled from) and the temperature provides a mechanism to yield highly diverse outcomes.

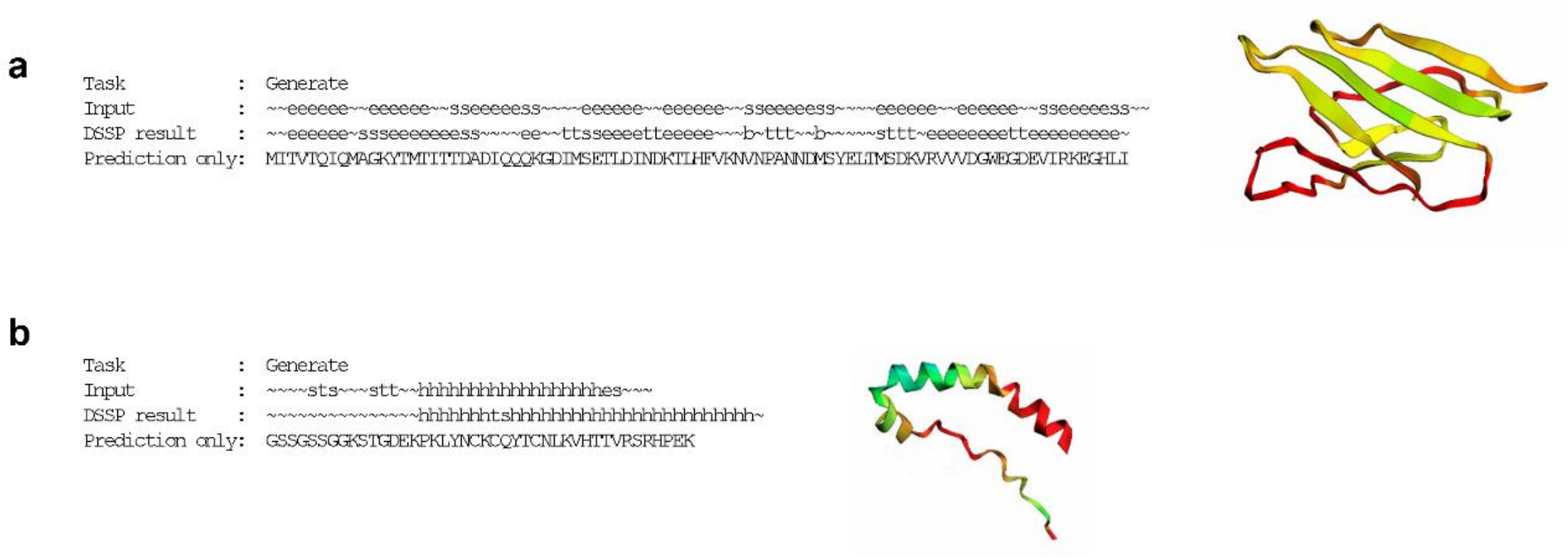

**Figure 8**: Panel **a** shows the design of a beta-sheet rich protein structure, using the prompt ~Generate<~~eeeeee~~eeeeee~~sseeeeess~~~~eeeeee~~eeeeee~~sseeeeess~~~~eeeeee~~eeeeee~~sseeeeess**~~>[MITVTQIQMAGKYTMTITTDADIQQQKGDIMSETLDINDKTLHFVKNVNPANNDMSYELTMSDKVRVVVDGWEGDEVIRKEGHLI]$.** Panel **b** shows a design task that yields a combination of a random coil and an alpha-helix. The validity of the predicted protein compared against the design task ("input") can be confirmed.

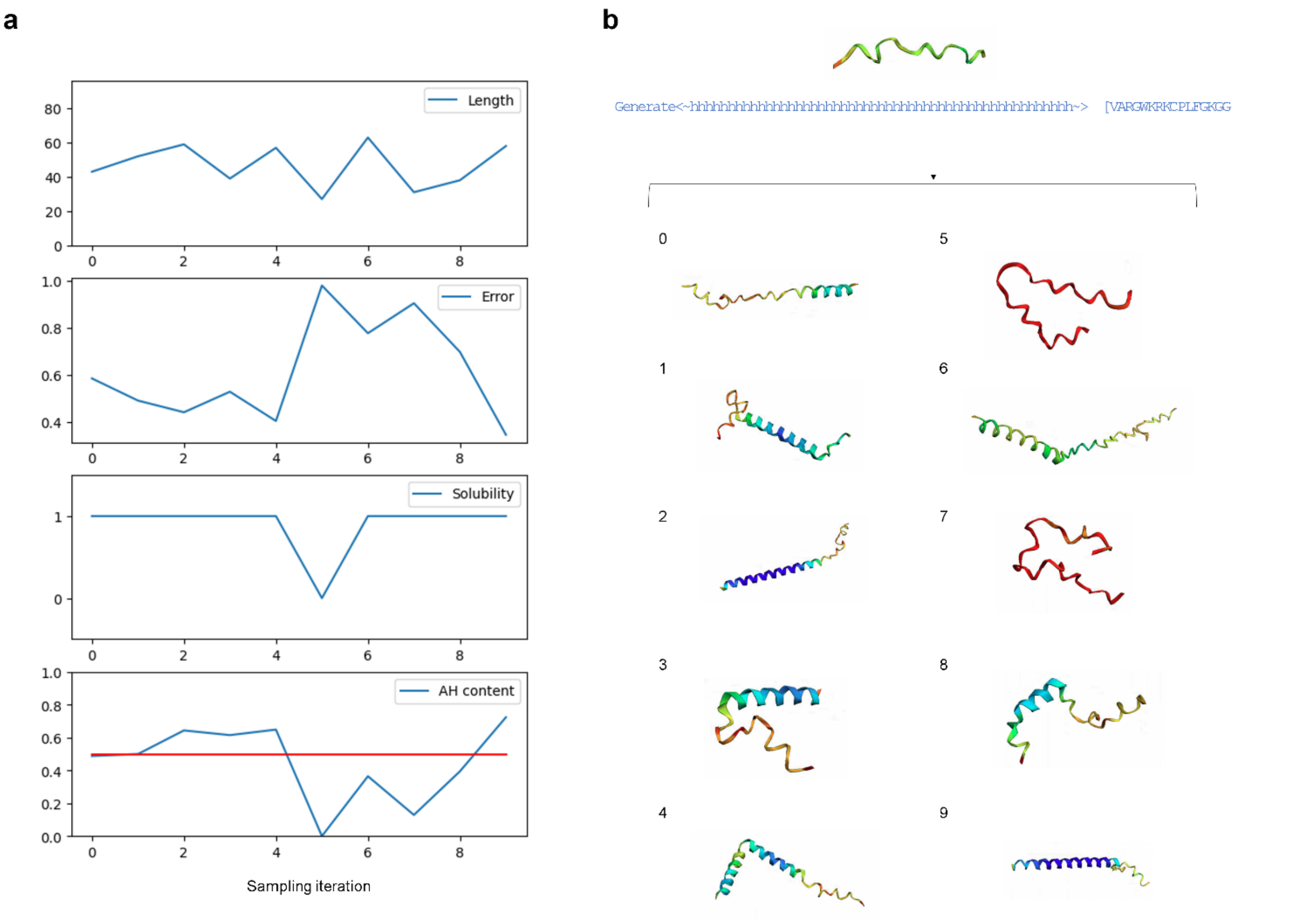

**Figure 9**: Example application in alpha-helical antimicrobial peptide design. Starting from 2MWL (amino acid sequence: `VARGWKRKCPLFGKGG`), an antimicrobial peptide (a new peptide design that shows antimicrobial activity against Gram-negative *E. coli* as well as plant pathogens, specifically *X. oryzae* and *X. campestris*). While the original peptide is unstructured, we seek to develop sequences that include the motif `VARGWKRKCPLFGKGG` but that yields a alpha-helix rich design that will likely help in assembly of structural materials and films. To do this we use the Generate task, assess designs against structural properties and solubility, and create a set of possible designs that can be screened for performance. Panel **a** shows an overview of the various metrics used. Panel **b** shows visual representations of the candidate proteins. The best performing candidate is sample number 9, being a peptide to be predicted to be soluble with the highest alpha-helix content. Sampling was conducted with *T*=.5, and filter_thres=0.9. This repeated sampling also shows that the model can reliably predict proteins at the desired length (the length cue is given by the secondary structure specification in the prompt: `Generate<~hhhhhhhhhhhhhhhhhhhhhhhhhhhhhhhhhhhhhhhhhhhhhhhhhhh~> [VARGWKRKCPLFGKGG`).

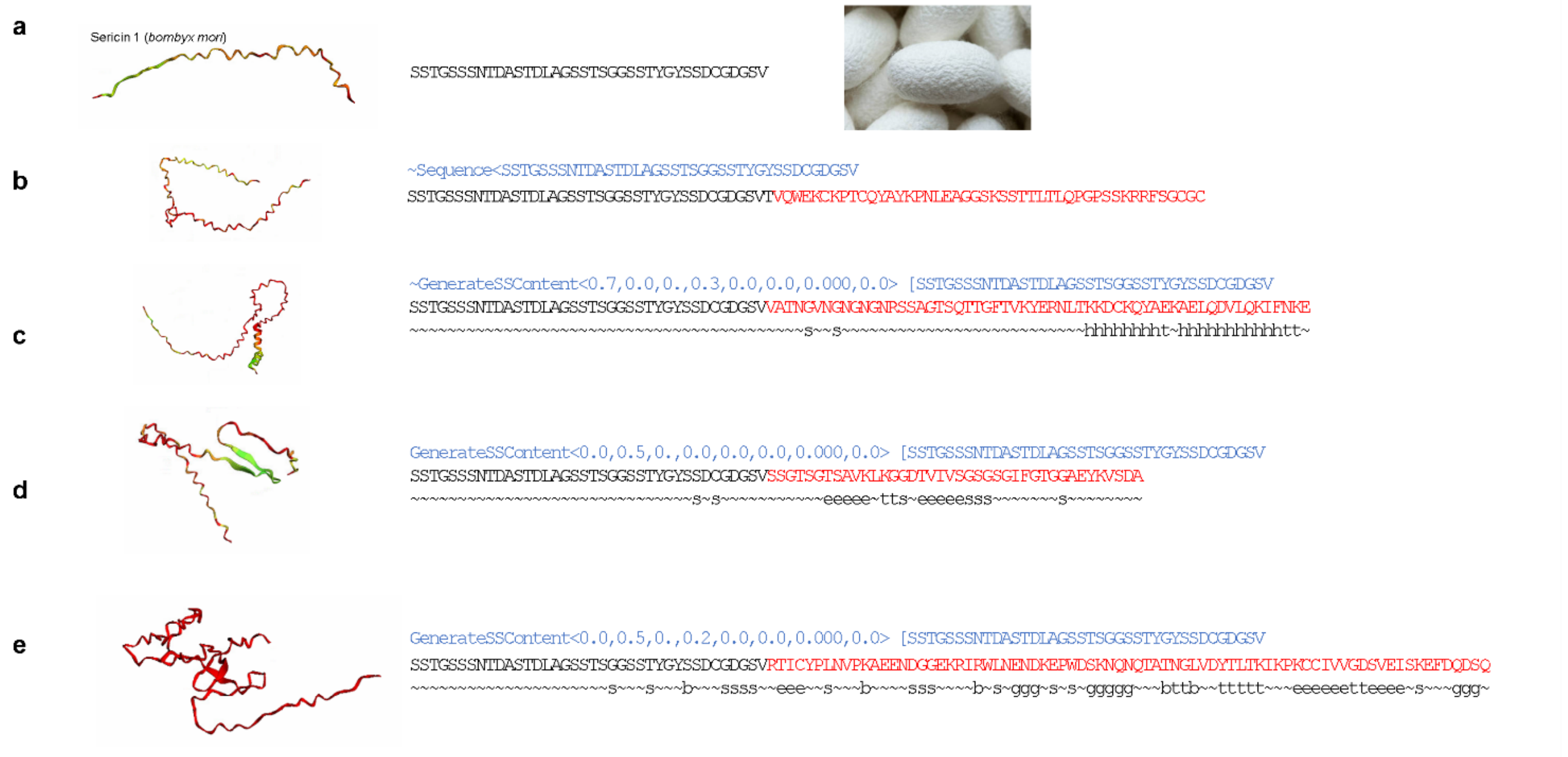

**Figure 10**: Using an amino acid sequence extracted from an existing protein, Sericin 1 (bombyx mori, P07856, SERI1_BOMMO, ser1 gene), and re-engineering the natural protein towards particular design objectives. Herein, panel **a** shows the original proteins structure and sequence of sericin. Panel **b** shows a sequence completion task, where the initial sequence is continued in an unconstrained manner. Panel **c** shows a design task where the design objective is provided alongside the original sequence and then continued to meet the design task. The design task in this case is to generate an alpha-helical protein, which is indeed found towards the end of the protein. Panels **d** shows a similar example, however, with the design task to generate a beta-sheet rich protein. This task is more difficult, but after a few trials a solution that meets the design target is obtained. Finally, panel **e** shows another example where the design task is given is a target with 50% beta-sheet, 20 random coil. This results in a more complex overall protein structure.

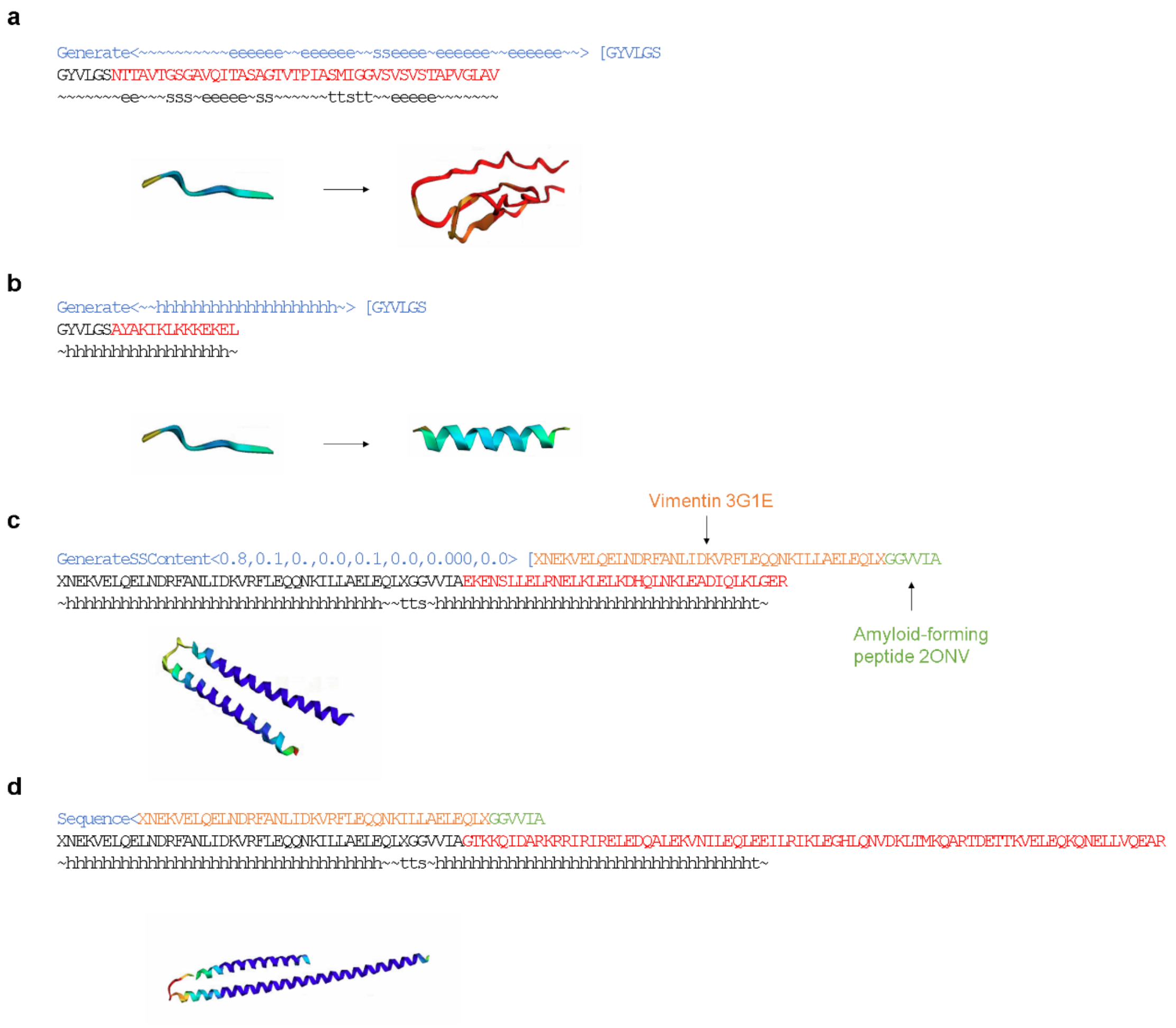

**Figure 11**: Experiments using a protein sequence designed earlier (see **Figure 5**, bottom example**)**, we expand it using residue-level secondary structure design. As can be seen in panel **a**, the random-coil sequence GYVLGS can be transformed into a beta-sheet rich structure. Similarly, it can be engineered to form an alpha-helix rich protein. Panels **c** and **d** show experiments where we use two naturally occurring proteins, vimentin 3GE1 and amyloid-forming peptide 2ONV, and query the algorithm to create an alpha-helix rich product. Similarly, in panel **d** we show an experiment where we use this combined sequence and query the algorithm to continue the sequence using the Sequence task. This results in an alpha-helix rich structure.

**Table 1**: Summary of DSSP secondary structure codes used in the modeling (the tables show both, DSSP 8 and DSSP 3 codes)

| **DSSP 8 code** | **Description** |
|---|---|
| h | Alpha-helix (AH) |
| e | Extended parallel and/or anti-parallel beta-sheet (BS) conformation |
| t | Hydrogen bonded turn (3, 4 or 5 turn) |
| ~ | Unstructured |
| b | Beta-bridge (single pair beta-sheet hydrogen bond formation) |
| g | 3/$3_{10}$ helix |
| i | pi-helix |
| s | Bend |
| **DSSP 3 code** | **Description** |
| h | Alpha-helix (AH) (h, g, i from DSSP 8) |
| e | Beta-sheet (BS) (b and e from DSSP 8) |
| ~ | Unstructured (~, t, s from DSSP 8) |

**Table 2:** Summary of all prompts used in the model. All tasks take the format Task<input_to_task> **[output].** During training, samples of the entire task and output is provided and trained using causal masking. During inference, we provide only the task input (starting with start token $\mathcal{T}$), and the model then solves the task by completing the prediction by providing the output, terminated with the end token $\mathcal{E}$ (start and end tokens not shown here for enhanced visual clarity).

| | **Task input** | **Description** | **Output example** |
|---|---|---|---|
| Pre training | Sequence<VFIYTDANGQV> | Used in pretraining, learn amino acid sequences | - |
| | SSSequence<hhhsseeeeeeeee~~~~~e> | Used in pretraining, learn secondary structure sequences (DSSP8) | - |
| Forward | Calculate< VFIYTDANGQ> | Calculate per-residue secondary structure (DSSP8) | **[~~~~~ee~hhhhttseetteeeee….]** |
| | CalculateSSContent<VFIYTDANGQV> | Calculate overall secondary structure content (8 ratios in DSSP8, can be converted into DSSP3 as per **Table 1**) | **[0.008,0.542,0.068,0.220,0.000, 0.000,0.000,0.161]** |
| | CalculateSolubility<VFIYTDANGQV> | Calculate solubility of protein sequence | **[1]** |
| Inverse | Generate<~hhhhhhhhhh~> | Generate amino acid sequence based on per-residue secondary structure | **[GLFILVLLLIVVAIFG]** |
| | GenerateSSContent<[0.008,0.542,0.068,0.220,0.000,0.000,0.000,0.161> | Generate amino acid sequence based on overall secondary structure content | **[KNKQGYAIPLVHCLQADVKFPV…]** |
| | GenerateSolubility<1> | Generate amino acid sequence based on overall solubility | **[VIENNVKYAVIENNVKYAQRDLQRDL…]** |